\documentclass[ngerman,english]{bvm} 

\bvmdef\type{P} % wird w{\"a}hrend der Nachbearbeitung und NICHT DURCH DIE AUTOREN eingetragen: Beitragsart: ``V'' = Vortrag, ``P'' = Poster, ``S'' = Softwaredemonstration

\bvmdef\articlenumber{3020}

\date{}
\title{Learning to be EXACT}
\titlerunning{Learning to be EXACT}

\subtitle{Cell Detection for Asthma on Partially Annotated Whole Slide Images}

\author{Christian~Marzahl$^{1,2}$,
Christof~A.~Bertram$^{3}$, 
Frauke~Wilm$^{1}$,
J{\"o}rn~Voigt$^{2}$,
Ann~K.~Barton$^{4}$, 
Robert~Klopfleisch$^{3}$, 
Katharina~Breininger$^{1}$, 
Andreas~Maier$^{1}$,
Marc~Aubreville$^{5}$}

\authorrunning{Marzahl et al.}

\institute{
$^{1}$Pattern Recognition Lab, Department of Computer Science, Friedrich-Alexander-Universit{\"a}t Erlangen-N{\"u}rnberg (FAU), Germany  \\
$^{2}$R \& D Projects, EUROIMMUN Medizinische Labordiagnostika AG \\
$^{3}$Institute of Veterinary Pathology, Freie Universit{\"a}t Berlin, Germany \\
$^{4}$Equine Clinic, Freie Universit{\"a}t Berlin, Berlin, Germany\\
$^{5}$Technische Hochschule Ingolstadt, Ingolstadt, Germany
}

\email{c.marzahl@euroimmun.de}

\begin{document}

%==============================================================================
% wählen Sie mit dem Befehl \selectlanguage die Sprache aus, in der Ihr 
% Proceeding verfasst ist
%
%\selectlanguage{german}
\selectlanguage{english}

\maketitle

\begin{abstract}

% Introduction
Asthma is a chronic inflammatory disorder of the lower respiratory tract and naturally occurs in humans and animals including horses. The annotation of an asthma microscopy whole slide image (WSI) is an extremely labour-intensive task due to the hundreds of thousands of cells per WSI.  
% Methods 
To overcome the limitation of annotating WSI incompletely, we developed a training pipeline which can train a deep learning-based object detection model with partially annotated WSIs and compensate class imbalances on the fly. With this approach we can freely sample from annotated WSIs areas and are not restricted to fully annotated extracted sub-images of the WSI as with classical approaches.
% Results 
We evaluated our pipeline in a cross-validation setup with a fixed training set using a dataset of six equine WSIs of which four are partially annotated and used for training, and two fully annotated WSI are used for validation and testing. 
Our WSI-based training approach outperformed classical sub-image-based training methods by up to 15\% $mAP$ and yielded human-like performance when compared to the annotations of ten trained pathologists.   
% Discussion
%Our evaluation and trained model is freely available at GitHub. 

\end{abstract}

\section{Introduction}
%

% Pathogenese?

Asthma is a chronic inflammatory disorder of the lower respiratory tract and can occur in multiple species. %Patients with asthma suffer from reversible airflow obstruction triggered by bronchoconstriction, airway edema, airway hyperresponsiveness or airway remodeling ~\cite{national2007expert}. 
%Asthma is subdivided into the five phenotypes intermittent, persistent, exercise-associated, aspirin-sensitive, or severe asthma ~\cite{national2007expert}. 
While asthma can affect humans, horses can also suffer from asthma and are often used as models for human disease \cite{3020-01} due to their similar symptoms and pathogenesis. The gold standard for diagnosis of equine and human asthma is to collect bronchoalveolar lavage fluid (BALF) and to examine the sample under a microscope or on digitised whole slide images (WSIs). Asthma and other pulmonary disorders are diagnosed based on the relative proportion of different cell types including eosinophils, mast cells, neutrophils, macrophages and lymphocytes. This typically requires manual counting of 300-500 cells and is therefore time-consuming and strenuous for the pathologist~\cite{3020-02}. Therefore, automatic solutions that support this task are of high interest. Although asthma is a common disease in horses and humans, to the author's knowledge there is no trained model or method published analysing asthma on WSI automatically.
For the development of machine learning algorithms, huge annotated datasets are generally required. A particular challenge for the annotation process of asthma is the large number of cells per WSI, which can easily reach hundreds of thousands. This makes the annotation process very labour-intensive and time-consuming. In order to increase the efficiency of the annotation process, expert-algorithm collaboration can be used where experts enhance pre-computed annotations of a trained model. Marzahl ~\etal\cite{3020-02} showed this for mitotic figures or pulmonary haemorrhage. An alternative option is to annotate multiple WSIs only partially to capture the domain variability between WSIs, and train a network to complete the annotation. On the one hand, training deep learning-based methods on partially annotated WSIs faces some additional challenges regarding tracking annotated WSI areas and leads to higher demands on the coordination and synchronisation between the participating institutes. On the other hand, training on partially annotated data simplifies training pipelines in terms of data augmentation and live patch sampling in contrast to extracted sub-image-based approaches. Nevertheless, sub-image-based approaches, where patches from the WSI have to be extracted before the training, are the only supported method for the most prominent object detection frameworks~\cite{3020-03} and are used in multiple WSI-based detection applications ~\cite{3020-04,3020-05}.  

%Also, for the development of new deep learning algorithms, the natural occurrence of asthma in horses has an enormous advantage as deep learning algorithms can be trained on equine data which can afterwards be applied to human samples. Less strict veterinary data privacy requirements support the creation of large-high-quality collections of datasets which are essential for the training of deep learning models. %To support this transferability of results between species, many data sets and methods in this area have been published in recent years ~\cite{bertram2019large}. 
% Probleme und Fragestellung
%Although asthma is a common disease in horses and humans , to the author's knowledge there is no trained model or method published to analyse asthma on WSI automatically. Furthermore, training deep learning-based methods on partially annotated WSIs faces some additional challenges regarding the tracking of which areas of the WSIs are already annotated but they also have some advantages regarding data augmentation in contrast to extracted sub-image based approaches. Nevertheless, sub-image-based approaches where patches from the WSI have to be extracted before the training, are the only supported method for the most prominent object detection frameworks~\cite{he2016deep} and are used in multiple WSI based detection applications ~\cite{kawazoe2018faster,meiquan2018cervical}. 
% Problems solution / contribution
As the main contribution to the field of deep learning-based cytological WSI analysis, we propose a training pipeline to train object detection models with live sampling on partially annotated WSIs.
Additionally, we create a baseline with a state-of-the-art deep learning-based object detection model for detecting five types of cells on WSIs. %Our best performing model is published and provides the groundwork for expert-algorithm collaboration on the additional WSIs. 
All code to train, evaluate, test our models and to reproduce our results for public is accessible at GitHub~\footnote[1]{https://github.com/ChristianMarzahl/Asthma\_WSI}. Furthermore, the WSIs can be accessed at reasonable request from the corresponding author.

%Thirdly, we developed an plugin for the open-source online platform EXACT\cite{marzahl2020exact} to qualify the cell distributions and put them into perspective with current grading recommendations for equine asthma. 

\section{Material and methods}

%Our research group collected a dataset 
The dataset consists of six cytological samples (Table \ref{3020-01}) of equine BALF which were cytocentrifugated and stained using May-Grunwald Giemsa stain. Afterwards, the glass slides were digitized using a linear scanner (Aperio ScanScope CS2, Leica Biosystems, Germany) at a magnification of 400$\times$ with a resolution of $ 0.25 ~\frac{\mu m}{px}$. Finally, two slides were completely annotated and the remaining four partially annotated by a veterinary pathologist.  %Most of the cell types are visually distinguishable from each other due to their distinct morphology but very closely packed and therefore difficult to separate. Finally, two slides were completely annotated and the remaining four partially annotated by a veterinary pathologist. 
%in a previous study \cite{10.1007/978-3-030-59710-8_3} 
Twenty patches from the same six WSI have been used in a recent study %[blinded for peer review]
~\cite{3020-02} to investigate the annotation accuracy from ten trained pathologist. We exclude these twenty patches from training to compare the accuracy of human experts with our algorithmic approach. 

\begin{figure}%[hbt!]
\centering
\includegraphics[height=0.4\textwidth]{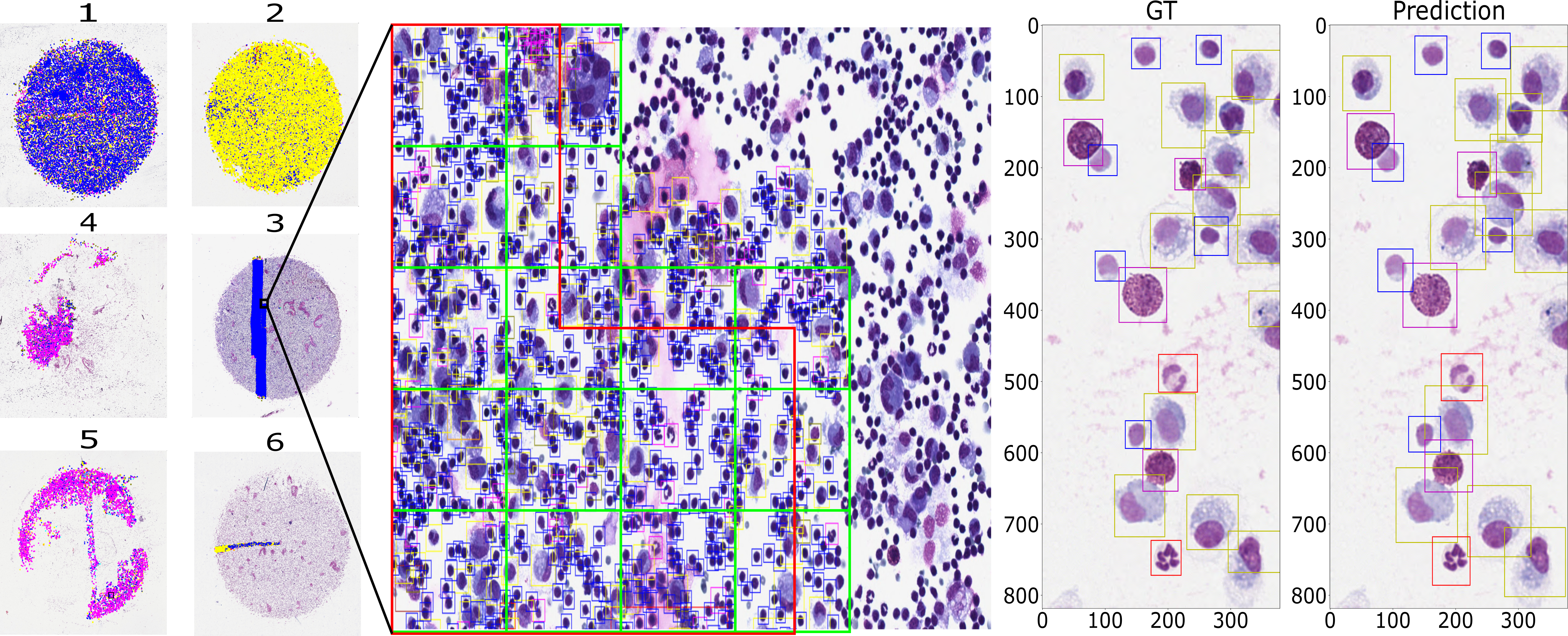}
\caption{Left: visualisation of the annotated regions/cells of the included slides. Center: visualisation of the 16 traditional fully annotated sub-images for sampling in green and 624 cell-based sampling positions for our live sampling approach within the red area. To prevent sampling from unannotated WSI areas, the red area is restricted to half of the patch size to the annotation border. Right: The ground truth (GT) and the deep learning based predictions; neutrophils (red), eosinophils (green), lymphocytes\~(blue), macrophages (yellow), mast cells (purple).}
\label{3020-fig1}
\end{figure}

\begin{table}[]
    %\centering
    \caption{Overview of the dataset including the file id, the number of cells per type and the screened sample area. The top two rows represent the completely annotated validation and test slides for the cross validation with a fixed train set.}
    \begin{tabular}{cccccccc}
    
     \hline
    id & eosinophils & mast cells & neutrophils & macrophages & lymphocytes & total & screened image  \\ \hline  

    1  &  21  &  511  &  3301  &  3934   &  14846  &  22613  &  100\%\\  %BAL Promyk Spray 4.svs
    2  &  47  &  762  &  951   &  16748  &  10342  &  28850  &  100\%\\ \hline  % BAL AIA Blickfang Luft.svs
    3  &  10  &  69   &  1321  &  3081   &  15666  &  20147  &  8\%\\ %  BAL 1 Spray 2.svs
    4  &  20  &  37   &  2467  &  729    &  2144   &  5397   &  28\%\\ % BAL Booker Spray 3.svs
    5  &  8   &  116  &  4491  &  1639   &  3077   &  9331   &  43\%\\  % BAL Bubi Spray 1.svs
    6  &  2   &  40   &  26    &  370    &  323    &  761    &  1\%\\ \hline  % BAL cent blue Luft 2.svs
       &  108 &  1535 &  12557 &  26501  &  46398  &  87099  &  46\%\\ \hline  

    \end{tabular}
    \label{3020-01}
\end{table}

\subsection{Label generation and training pipeline}
The dataset containing only six WSIs appears to be comparably small. However, the cells in the WSIs are annotated by experts and subdivided into five classes using SlideRunner~\cite{3020-06}, rendering it one of the largest manually annotated cytology datasets to date.
%Although six WSIs appears to be a comparably small dataset, its with SlideRunner\cite{aubreville2018sliderunner} 87099 expert-annotated cells subdivided into five classes, rendering it one of the largest manually annotated cytology datasets to date.
The dataset displays an extreme class imbalance, with the rarest class of eosinophils representing only 0.12\% of all annotations.
A particular challenge for training neuronal networks emerges from the sparse annotation of four WSIs, as shown in the column "screened" in table \ref{3020-01}. %It is not unusual that the training of neural networks is started while the datasets is still incomplete. This offers the possibility to speed up the annotation of the dataset and to increase the quality of the data by means of expert-algorithm cooperation \cite{10.1007/978-3-030-59710-8_3}, but also leads to higher demands on the coordination and synchronisation between the participating institutes. 
To meet this challenge of partially annotated WSIs, we apply an online training approach using the open-source online annotation platform EXACT~\cite{3020-07}. EXACT supports a persistent screening mode that allows experts to screen WSIs in a self-determined resolution. By reusing the information this screening mode provides, we are able to track exactly which areas of the slide have already been annotated by the expert and can download them into the training process via the provided REST-API. 
%Furthermore, rapid single click annotations for bounding boxes are supported, which incorporate default sizes for cell types to reduce the need for manual corrections. This is especially true for asthma cells where the cell types vary in their default size and therefore a pre-selected annotation size per type reduces the need for manual corrections. Also, EXACT offers an asynchronous REST-API, which allows for example downloading or uploading slides, annotations, and the progress of the WSI screening independently from the programming languages or frameworks.

%\subsection{Training pipeline}

%EXACT eliminates the need to send data records and annotations by e-mail or hard disk between the participating institutes which is the case for offline tools like SlideRunner \cite{aubreville2018sliderunner} or QuPath \cite{bankhead2017qupath}. 
We propose the following online training pipeline for equine asthma. 
As a first step, screened areas of the slides and associated annotations are downloaded from the EXACT server via the REST-API. %EXACT's annotation versioning system allows to decide which annotation state is downloaded, for example, an old state for reproducing results or the current state. 
The training pipeline is then initialised with the downloaded information, the network architecture and the loss function. 
During training, new patches are sampled live from the WSI according to the patch selection and sampling strategy described in the following sections and are restricted by information about the screened area provided by EXACT.
%During training, the training pipeline allows to continuously integrate new annotations in the training by synchronising the train, validation and test data accordingly. 
The trained models can be applied to new data and the results can be synchronised with the server for expert review. 

\subsection{Live patch selection and sampling strategy}

%The sampled WSI patches for training are of size 1024$\times$1024\,px. This large patch size for deep learning is motivated by the size of immune cells characteristic for asthma, which ranges from around 25$\times$25\,px for lymphocytes up to 240$\times$240\,px for macrophages in our dataset. This varying cell size has multiple consequences. To obtain more favourable detection performance for larger cells, we have to ensure that a sampled patch covers the complete cell. This can be done by sampling with a lower resolution, but this would further reduce the size of lymphocytes and therefore decrease their accuracy. Alternatively, increasing the patch sizes is limited by graphic cards memory. Therefore, a patch size of 1024$\times$1024\,px represents a compromise between high accuracy and sufficient batch sizes.

To counteract the described class imbalance and partial annotations, we propose the following sampling strategy which uses annotated cells as seeds for training patches: Each training patch has a cell that was manually annotated by an expert in its center. Consequently, only cells that have at least a distance of half of the patch size to the border of the annotated region can be used as patch centers (red area in Figure \ref{3020-fig1}). Each training batch contains at least five patches with each of the five cell types represented as patch center cell. The center cells within each class are randomly chosen from the annotations. If the training batch size is chosen larger than the number of cell types, a new cell type is randomly selected with a probability proportional to $1 - p_k$ where $p_k$ is the relative class frequency of each cell type until the required batch size is reached. This results in a sampling strategy which can freely choose the sampled patches and reduces the possibility of sampling a given region repeatedly. This is highly desirable to counter overfitting and works as an advanced augmentation technique.

For comparison we extract all available fully annotated areas as sub-images of the WSIs to simulate a traditional training pipeline. This results in a total of 1862 sub-images of which 851 belong to the two fully annotated test WSIs. Example sub-images are visualised in Figure \ref{3020-fig1} on the right with green rectangles. For training the same cell type-based sampling strategy to counteract the described class imbalance is applied. 

\subsection{Object detection methods}

%One deep learning-based object detection approach which showed remarkable results on cell detection tasks~\cite{marzahl2020deep,bertram2019large} is RetinaNet~\cite{lin2017focal}. Cell detection can be approached as an object detection task~\cite{marzahl2020deep,liang2016cnn}.

Since the training strategy itself is the main contribution of this work, we use a publicly available and for cytology optimised implementation~\cite{3020-08} of the successful RetinaNet~\cite{3020-09} architecture.
Different ResNet-variants \cite{3020-10} (ResNet-18, -34, -50) pretrained on ImageNet%~\cite{deng2009imagenet} 
are applied as backbone networks with appropriate mini-batch sizes. The networks are trained using the sub-images-based and the proposed live sampling-based approach with a patch size of 1024$\times$1024\,px. Each of the two fully annotated WSIs (Table \ref{3020-01}) are used once as the validation set and once as the test set while keeping the training set static to allow for a form of cross-validation given the limited amount of cases. During training, data augmentation (rotation between zero and 90 degrees, horizontal and vertical flips, random increase or decrease of intensity in the range of -20 to +20\%) is applied and the networks are trained until convergence on the validation set. The initial learning rate is set to 1e-3 and reduced to 1e-4 and 1e-5 if the validation loss doesn't decrease for three epochs. One epoch consists of 500 training patches regardless of the WSI-based sampling mode or the extracted sub-images. The object detection accuracy is measured as mean Average Precision (mAP) according to the 2007 PASCAL VOC challenge. %~\cite{everingham2010pascal}

%\subsection{EXACT asthma plugin and inference}
%The microscopy-based clinical evaluation of asthma is determined by the relative relationship between the different cell types. In order to support the user in the analysis of his findings, we have developed an EXACT plugin which calculates the relative proportions for the relevant cells and displays them clearly in a tabular form together with the clinical threshold values and their source. The plugin also supports a deep learning-based inference of our asthma model to the current section of the slide. Afterwards, the user can confirm or improve the models' predictions. The model is JavaScript-based executed in the browser. 

\section{Results}

\begin{table}[]
    \centering
    \caption{The mean average precision for the five types of cells in respect to the number of layers used for the ResNet backbone network (BB) and batch size (BS). The modes represent our method working on partially annotated WSIs and a classical approach with extracted sub-images.}
    \begin{tabular}{cccccccccc}
     \hline
    mode & BB & BS & eosinophils & mast cell & neutrophils & macrophages & lymphocytes & $\varnothing$	 \\ \hline  
    
    ours  & 18 & 16 & 0.93 & 0.85 & 0.88 & 0.89 & 0.81 & 0.87 \\
    %sub-image & 18 & 16 & 0.10 & 0.48 & 0.68 & 0.80 & 0.58 & 0.53 \\ No strategy
    sub-image & 18 & 16 & 0.69 & 0.72 & 0.68 & 0.80 & 0.71 & 0.72 \\
    
    ours  & 34 & 16 & 0.91 & 0.86 & 0.90 & 0.89 & 0.78 & 0.87 \\
    %sub-image & 34 & 16 & 0.10 & 0.48 & 0.68 & 0.81 & 0.57 & 0.53 \\ No strategy
    sub-image & 34 & 16 & 0.70 & 0.71 & 0.68 & 0.80 & 0.72 & 0.72 \\
    
    ours  & 50 & 6  & 0.92 & 0.80 & 0.90 & 0.89 & 0.81 & 0.86 \\  
    %sub-image & 50 & 6  & 0.11 & 0.47 & 0.68 & 0.81 & 0.56 & 0.52 \\ No strategy
    sub-image & 50 & 6 & 0.72 & 0.69 & 0.68 & 0.81 & 0.75 & 0.73 \\ \hline  
    
    %1 & our   & 18 & 16 & 0.93 & 0.81 & 0.91 & 0.81 & 0.74 & 0.84 \\
    %1 & image & 18 & 16 & 0.20 & 0.52 & 0.66 & 0.77 & 0.57 & 0.54 \\
    
    %1 & our   & 34 & 16 & 0.94 & 0.81 & 0.91 & 0.81 & 0.74 & 0.85 \\
    %1 & image & 34 & 16 & 0.20 & 0.51 & 0.66 & 0.78 & 0.55 & 0.54 \\
    
    %1 & our   & 50 & 6  & 0.93 & 0.74 & 0.91 & 0.84 & 0.75 & 0.83 \\  
    %1 & image & 50 & 6 & 0.19 & 0.50 & 0.66 & 0.78 & 0.54 & 0.53 \\
    %\hline \hline 
    %2 & our   & 18 & 16 & 0.92 & 0.89 & 0.85 & 0.96 & 0.88 & 0.90 \\
    %2 & image & 18 & 16 & 0.00 & 0.43 & 0.70 & 0.82 & 0.59 & 0.51 \\
    
    %2 & our   & 34 & 16 & 0.88 & 0.91 & 0.89 & 0.96 & 0.82 & 0.89 \\
    %2 & image & 34 & 16 & 0.00 & 0.45 & 0.70 & 0.83 & 0.58 & 0.51 \\
    
    %2 & our   & 50 & 6  & 0.90 & 0.85 & 0.89 & 0.94 & 0.86 & 0.89 \\ 
    %2 & image & 50 & 6  & 0.03 & 0.44 & 0.69 & 0.84 & 0.57 & 0.51 \\

    \end{tabular}
    \label{3020-02}
\end{table}

Independent of the backbone model or batch size, the accuracy of our WSI-based sampling approach converge at an mAP of 0.87 (min=0.86, max=0.87, IoU=0.5, epochs=73), outperforming the model trained on sub-images (mAP=0.72, min = 0.72, max=0.73, IoU=0.5, epochs=23) as shown in Table \ref{3020-02}. After 23 epochs the sub-images-based training is terminated due to overfitting. The backbone network of the model has no effect on the accuracy. The live sampling approaches show the lowest performance for the lymphocytes, which are the smallest and most clustered type of cells. The sub-images-based approach scores lowest on the rare classes of eosinophils and mast cells due to overfitting.
%While the sub-images-based approach scored lowest on the rare classes of eosinophils and mast cells due to overfitting, we attribute this overfitting to the sub-images-based method which can only sample from the pre-extracted images. In contrast, the live approach can almost sample from the complete screened area, which massively reduces the chance to sample the same patch twice and therefore mitigate overfitting. 

When applying the trained solution with a ResNet-18 as a backbone on the image patches and ground-truth published in [blinded for peer review]% ~\cite{10.1007/978-3-030-59710-8_3}
, we reach a mean mAP across images of 0.76 with the proposed method and 0.63 with the sub-images-based approach compared to the experts reaching a published mean concordance of $\mu$=0.73 mAP (min=0.56, max=0.82,$\sigma$=0.08).

%Compared to the study on twenty asthma patches~\cite{10.1007/978-3-030-59710-8_3}, where ten pathology experts reached a mean concordance with the published ground truth of $\mu$=0.73 mAP (min=0.56, max=0.82,$\sigma$=0.08), our WSI-based sampling method reached a mean concordance of $\mu$=0.76 mAP exceeding the sub-images-based approach with $\mu$=0.63 mAP. 

\section{Discussion and outlook}

We demonstrated the creation of an object detection training pipeline which is able to use partially annotated WSIs efficiently and is superior to a simple sub-image-based approach. Our proposed approach allows for better sampling strategies and data augmentation for rare classes  which massively reduces the chance to sample the same patch repeatedly and therefore mitigate overfitting, as apparent in the considerable difference in performance for eosinophils in the evaluation. This resulted in a object detection model with human like performance on a small set of example patches. However, this work has the limitation that only six images have been partially annotated by one expert, which needs to be addressed in further research. This work can be used as a baseline for further enhancements, like increasing the detection performance of small cells, optimising the non-maximum suppression algorithm for high quantities of cells but also to create new annotations in an expert-algorithm based manner on the remaining WSIs.
%which currently takes longer than the inference time

% The non-maximum suppression algorithm for high quantities of cells takes longer than the inference time
%But even at this early development stage the accuracy showed no significant difference to trained pathologist. all samples are collected from one scanner,

% Cons
% Mehr daten 
% Nur ein Model
% unvollständig gelabelt
% ein Scanner 
% kleine Zellen liefern schlechtere Ergebnisse
% WSI blöd

% Pro
% Mehr daten verfügbar
% Von daten bis inference für jederman
% Grundlage für die Zukunft. Modelle leicht zu intergrieren

\ack{CAB gratefully acknowledges financial support received from the Dres. Jutta \& Georg Bruns-Stiftung f\"ur innovative Veterin\"armedizin.}

\bibliographystyle{bvm}

\bibliography{3020}

\begin{thebibliography}{10}

\bibitem{3020-01}
Bullone M, Lavoie JP.
\newblock Asthma “of horses and men”—how can equine heaves help us better
  understand human asthma immunopathology and its functional consequences?
\newblock Mol Immunol. 2015;66(1):97--105.

\bibitem{3020-02}
Marzahl C, Bertram CA, Aubreville M, et~al.
\newblock Are Fast Labeling Methods Reliable? A Case Study of Computer-Aided
  Expert Annotations on Microscopy Slides.
\newblock In: MICCAI. Cham: Springer International Publishing; 2020.  p.
  24--32.

\bibitem{3020-03}
Huang J, Rathod V, Sun C, et~al.
\newblock Speed/accuracy trade-offs for modern convolutional object detectors.
\newblock In: CVPR; 2017.  p. 7310--7311.

\bibitem{3020-04}
Kawazoe Y, Shimamoto K, Yamaguchi R, et~al.
\newblock Faster r-cnn-based glomerular detection in multistained human whole
  slide images.
\newblock Imaging. 2018;4(7):91.

\bibitem{3020-05}
Yang F, Yu H, Silamut K, et~al.
\newblock Parasite Detection in Thick Blood Smears Based on Customized
  Faster-RCNN on Smartphones.
\newblock In: AIPR; 2019.  p. 1--4.

\bibitem{3020-06}
Aubreville M, Bertram C, Klopfleisch R, et~al.
\newblock SlideRunner.
\newblock In: Bildverarbeitung f{\"u}r die Medizin 2018. Springer; 2018.  p.
  309--314.

\bibitem{3020-07}
Marzahl C, Aubreville M, Bertram CA, et~al.
\newblock EXACT: A collaboration toolset for algorithm-aided annotation of
  almost everything.
\newblock arXiv preprint arXiv:200414595. 2020;.

\bibitem{3020-08}
Marzahl C, Aubreville M, Bertram CA, et~al.
\newblock Deep Learning-Based Quantification of Pulmonary Hemosiderophages in
  Cytology Slides.
\newblock Sci Rep. 2020;10(1):1--10.

\bibitem{3020-09}
Lin TY, Goyal P, Girshick R, et~al.
\newblock Focal loss for dense object detection.
\newblock In: Proceedings of the IEEE international conference on computer
  vision; 2017.  p. 2980--2988.

\bibitem{3020-10}
He K, Zhang X, Ren S, et~al.
\newblock Deep residual learning for image recognition.
\newblock In: CVPR; 2016.  p. 770--778.

\end{thebibliography}
% Bitte setzen Sie hier Ihre Beitragsnummer ein und benennen Sie
% die BibTeX-Datei ebenfalls auf Ihre Beitragsnummer um.
%Kontrollzeiledef
\marginpar{\color{white}E\articlenumber} % Zeile nicht verändern!
\end{document}